# HANDLING TRUST IN A CLOUD BASED MULTI AGENT SYSTEM

Imen Bouabdallah[1] and Hakima Mellah[2]

[1]Department of Computer Science, USTHB, Bab ezzouar, Algeria
[2]Information and multimedia system Department,
CERIST, Ben Aknoun, Algeria

## ABSTRACT

*Cloud computing is an opened and distributed network that guarantees access to a large amount of data and IT infrastructure at several levels (software, hardware...). With the increase demand, handling clients' needs is getting increasingly challenging. Responding to all requesting clients could lead to security breaches, and since it is the provider's responsibility to secure not only the offered cloud services but also the data, it is important to ensure clients reliability. Although filtering clients in the cloud is not so common, it is required to assure cloud safety.*
*In this paper, by implementing multi agent systems in the cloud to handle interactions for the providers, trust is introduced at agent level to filtrate the clients asking for services by using Particle Swarm Optimization and acquaintance knowledge to determine malicious and untrustworthy clients. The selection depends on previous knowledge and overall rating of trusted peers. The conducted experiments show that the model outputs relevant results, and even with a small number of peers, the framework is able to converge to the best solution. The model presented in this paper is a part of ongoing work to adapt interactions in the cloud.*

## KEYWORDS

*Multi agent system, cloud, trust, interaction, PSO.*

## 1. INTRODUCTION

Due to the pandemic of Covid-19 and the increase of remote work, the use of cloud computing has been highly increased [1] [2], leading to a higher number of both cloud clients and cloud providers.

The high number of users can cause the emergence of malicious entities Identifying malicious agents between the large numbers all at once would be difficult due to the insufficient amount of information regarding the users. The non-detection of such entities can produce security problems and restrictions in cloud use [3].

To overcome the extant security flaws in the cloud many paradigms have been introduced to improve the protection levels or reinforcing security.

Multi agent system (MAS) is a distributed structure [4] that has been employed to solve complex problems such us detecting security flaws in multiple fields.

MAS include multiple interacting autonomous agents existing in a shared environment. The agents are distributed all over the system and capable of remaining connected through their interactions mechanism. Their intelligence allows them to autonomously take actions in order to solve their assigned tasks and evolve in the system.

Since distribution is common in Multi Agent System (MAS) and the cloud, and both paradigms consist of interconnected agents (in MAS) or users (in the cloud), combining them would draw the strong aspects of both.

Yet, due to the openness of the cloud [5] (multiple entities exist and can leave or enter the network randomly) and diversity of users, the combination still present some flaws that was the reason for the lack of early research in this area.

Nowadays, MAS is explored through its strong features to respond and overcome the cloud's flaws. Multi Agent System features include autonomy, perception and awareness, reactivity, interaction, cooperation [5], and many more depending on the implementation. This work is a prelude for an adaptive interaction framework, the propose model uses interaction along with cooperation to share knowledge among agents to determine trustworthiness of cloud consumers.

In this paper, Multi Agent System interaction mechanisms are used along with Particle Swarm Optimization (PSO) in the cloud to determine the trustworthiness of the clients and eliminate the malicious users. The following section discusses some of the recent related works, section 3 presents a background to: define MAS and the cloud, introduce trust in the cloud and present the main actors of the system; section 4 discusses the proposed PSO-Trust model and section 6 presents the evaluation and experimentation results.

## 2. RELATED WORKS

### 2.1. Multi agent system across the cloud

MAS have successfully solved many problems in the cloud, and it is even more promising when combined with other promising techniques. E.g. enhancing security by detecting intrusions [6], decision-making, service discovery [7], service reservation [8] …

Implementing MAS in the cloud is using the agents of the system to represent the actors of the cloud system, execute their commands and negotiate instead of them.

Service discovery was studied in [9]; authors used a hybridization of MAS and web services. Where they presented sets of agents that act on behalf of cloud consumers and cloud providers to generate and agree to the contract, the service discovery agent acts on behalf of the client using semantic ontology to select the best-fitted service provider.

Despite the diverse works joining MAS with the cloud, the majority focus on the concept of agents' autonomy, but yet neglect a critical aspect: cooperation.

In this work, cooperation between agents is considered through knowledge sharing among acquaintance agents to provide more accurate data to the system. We make use of the interaction mechanisms of MAS to exchange relevant and up to date data.

## 2.2. Implementing trust in the cloud

Trust management in the cloud have been increasingly studied to improve both security and intrusion detection, and is employed for service discovery and recommendation.

In the cloud, privacy is a major concern for all users and providers. Most of the works considering trust in the cloud, employ it on the client side ([10] [11] [12]) to help choose a service provider.

Using users' feedback and behaviour in [11], they proposed an Evidence Trust model to select the trusted cloud service provider (CSP). The final assigned trust value is computed through cumulating feedback and behavioural trust and would determine the trustworthiness of a CSP. To detect false feedback they compared the previously submitted feedback from user with the current one. In this case, if the client is always giving a false feedback it will pass undetected.

Li, et al in [13] used trust along reputation in an IOT cloud-based to determine the trustworthiness of a cloud service, by using feedback rating of costumers and evaluation of security metrics.

Yet, the works on guaranteeing only trusted clients in the network are very insignificant; but since it's the provider's responsibility to secure the cloud, it is highly important to check clients' background before granting the service and experiencing the security problems (identity theft, data breach…), which this paper discuss.

For the reason that if a provider in the cloud is under attack or experiencing security flaws that would certainly influence clients data, but also provider's reputation which can hardly be changed, and may result in losing all users.

Mehraj and Banday discus in [14] an application of trust in cloud deployment using Zero-Trust engine where, instead of trust being handled by a role agent, they use the engine to determine trustworthy entities. As for end users, they relay on authentication to validate users' identity and therefore trustworthiness, which may not always be the source of user's credibility.

A dynamic multi-dimensional trust evaluation method was proposed in [15] to determine trustworthiness of cloud providers and cloud consumers. In this work, the trustworthiness is considered to define the honest behaviour of the consumer, and since it does not interact only with providers but also with: brokers, auditors and SLA agents, they collect compliance information from all entities to calculate trust.

## 2.3. Implementing trust using PSO

Bio-inspired algorithms follow real life principals and are evolutionary aiming to improve the quality of problem solving. They derive from the behaviour of large groups but still can be applied to small data and provide precise and accurate data [16].

Particle swarm optimization is based on social behaviour [17] and thus can make use of shared knowledge between entities of the system. It products precise results and does not require large data (compared to other algorithms).

For group decision making, [18] discuss how, by introducing trust, individuals in a group can be influenced by others to change their opinions and decisions. Individuals' opinions (or expert evaluations as given in the illustrating example) represent the particles of PSO that is employed to determine their optimum.

Authors argue, by comparing the proposed model to another without trust, that the drawbacks can be avoided using trust and making use of group intelligence.

In [19] PSO was introduced to optimize neural network (NN) parameters and use NN to determine costumers trust rate in the cloud, Although the experiments showed that the results were promising, the NN require large data and need training over multiple times to really outputs precise data. Unlike that, PSO-Trust model presented in this paper requires only small swarm population to converge to the best solution.

## 3. BACKGROUND

Multi agent systems and the cloud are the main technologies involved in this paper, the following section present a brief review of these paradigms.

### 3.1. Multi agent system
-
Multi agent system (MAS) [20] consists of multiple interacting intelligent agents that are capable of handling tasks and cooperating to solve complex problems. Each agent receives a different task and outputs a different action, that would eventually be collected to solve a complex problem representing the global goal of the system.

Intelligent Agents are autonomous and situated in an environment [21], they are capable of perceiving their surroundings, making and choosing their own decisions, and taking actions without assistance.

Agents are endowed of multiple features: autonomy, intelligence, interactivity, interaction… [21][5].

The latter is a mean of communication and cooperation; by interacting, agents would share knowledge or resources and would cooperate to accomplish complex tasks. Interaction's type changes with the goals and tasks. Negotiation, a type of interaction, is usually set up to settle over an agreement or to reach a compromise to outcome conflicts.

Agents' autonomy drew much attention to them in research area, and are therefore applied in various domain.

### 3.2. The cloud

Cloud computing [22] is a model for enabling ubiquitous, convenient, on-demand network access to a shared pool of configurable computing resources.

The users in the cloud are distributed over the network yet interconnected and able to access the needed resources at any time remotely.

These users can be classified, NIST [22] presents an architecture of five main roles in the cloud. In our opinion, and while implementing MAS along the cloud, the roles could be brought to two main classes: users and helper agents. In this work, we only consider users; however, the presence or absence of helper agents would not effect the model.

As for users, two actors are identified: cloud clients (consumers): the users asking for services, and cloud service providers: the actors offering the services at a given price with some defined criteria.

Many security flaws are present at this level effecting clients' privacy and data, and providers database and reputation [3]. To encounter some of the difficulties, multi agent systems (MAS) are introduced to make use of agents' intelligence for handling complex tasks. Agents would negotiate for the provider and study the opponent before and during the iteration.

Agent's interaction strategy would have a direct impact on data security and user satisfaction that can be gained by ensuring the protection of client's privacy. Moreover, for that we introduced trust.

### 3.2.1. Trust in the cloud

In literature, trust is defined as the firm belief in the reliability of the received information or also as "the vesting of confidence in persons or abstract systems, made on the basis of a leap of faith which brackets ignorance or lack of information" [23]. Between agent, trust could be defined based on multiple criteria that differentiate based upon the context which might lead to having different definition; in some works, it is based on reputation from other agents [24], reliability of the agent, respect of the given information, insurance of privacy…

Trust and security are constantly associated, mostly because trust engendera feeling of safety that is earned, acquired or build on strong knowledge [25]. Trusting an entity implies asserting the credibility of information it shares and the identity it presents and thus guarantying its access to data or platforms, which is major security concern.

### 3.2.2. Actors in the cloud based MAS

In order to clarify the subsequent sections, we start here by defining the main actors of the cloud-based system.

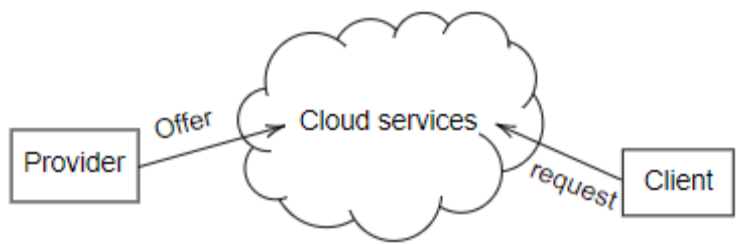

Figure 1. Main actors in the cloud

The cloud is a meeting environment for providers and clients (Figure 1), where the providers try to satisfy the clients need and expectations for services' quality.

Whereas clients (consumers) ask for services (software, database, collaborative environment…), providers work to deliver the best services at best cost and quality:

1. A cloud provider: delivers cloud computing services and solutions, responsible for making a service available for consumers [22],

2. A cloud consumer: could be individual or organization, using cloud services and resources, and maintain a business relationship with the provider. The consumer itself could represent a cloud provider.

In a cloud-based MAS, agents take actions for the cloud actors; client are represented by a client agent and providers by service agents (to handle the large amount of requests, multiple service agents are assigned to a single cloud provider (Figure 2)).

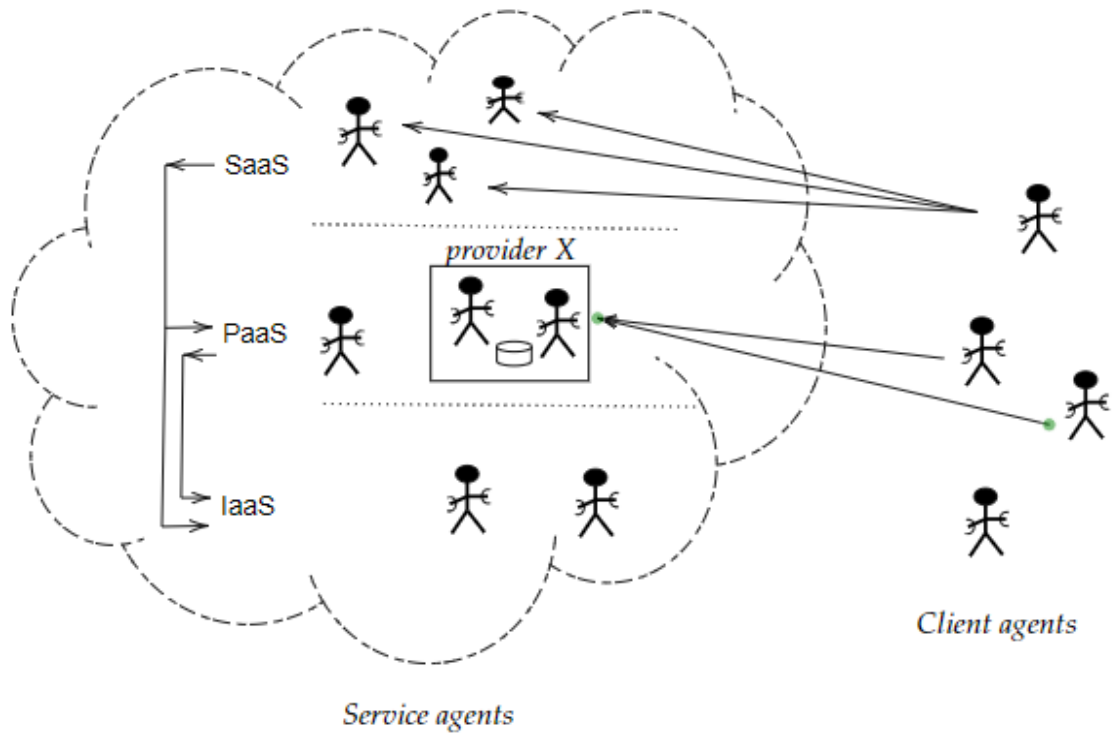

Figure 2. MAS cloud agents

- Client Agent (CA): carries cloud consumer request and handle the search and negotiation until service delivery,
- Service Agent (SA): represents the cloud service provider, it handles the consumers requests, it can also act as a cloud consumer to ask for complimentary services from other providers.

Agents conduct interactions instead of cloud actors to save time and effort because of agent's autonomy they are able to complete their assigned tasks without assistance.

For a service agent (SA: agent that guarantees services access) a client is trusted if it: respects the agreements, does not violate the cloud resources term of use and privacy, does not barging for so long that it cause famine (if the system does not have a timeout limit). On the other hand, a client could be less trusted if it has a bad reputation regarding the previous discussed points.

## 4. PSO-TRUST MODEL

In the following section, we will present the proposed model starting by defining PSO.

## 4.1. Particle Swarm Optimization

Particle Swarm Optimization (PSO) [16] is a search algorithm [26], inspired by the social behaviour of bird flocks. Each individual in PSO represents a potential solution and is called a particle. The whole population is referred to as a swarm.

The algorithm aims to find the best solution (global best) from the swarm by running multiple iterations until the whole population converges or the maximum number of iterations is exceeded. Each particle possesses a position, and a velocity that represents the speed and direction by which the particle moves at each iteration.

At each iteration, velocity and position are updated through the following equations:

$$v(t+1) = v(t) + c_1 r_1(t) \left(p^{best}(t) - x(t)\right) + c_2 r_2(t) \left(g^{best}(t) - x(t)\right) \quad (1)$$

$$x(t+1) = x(t) + v(t+1) \quad (2)$$

Where:

- t : time/iteration
- v(t) : velocity at time stamp t
- x(t) : position
- $c_1, c_2$ : constants for nostalgia and envy (resp.), also called trust parameters
- $r_1, r_2$ : random vectors $\in [0,1]$
- $p^{best}, g^{best}$: personnel and global best (resp.)

When a particle moves to a new position, it is effected by three component: previous position, social component and cognitive component.

The whole population's (swarm) size effects the convergence of the algorithm for: the smaller the size the slower to converge, and the bigger the size the more space to explore.

## 4.2. Model flowchart

The following model (Figure 3) casts a presentation of the proposed PSO-Trust model.

Upon receiving a service request, and instead of start negotiation on the spot, the SA would first start by checking the client's trustworthiness by either retrieving its trust weight from the database (if the agent have had previous interactions with the client), or acquire it from its acquaintances (Figure 3).

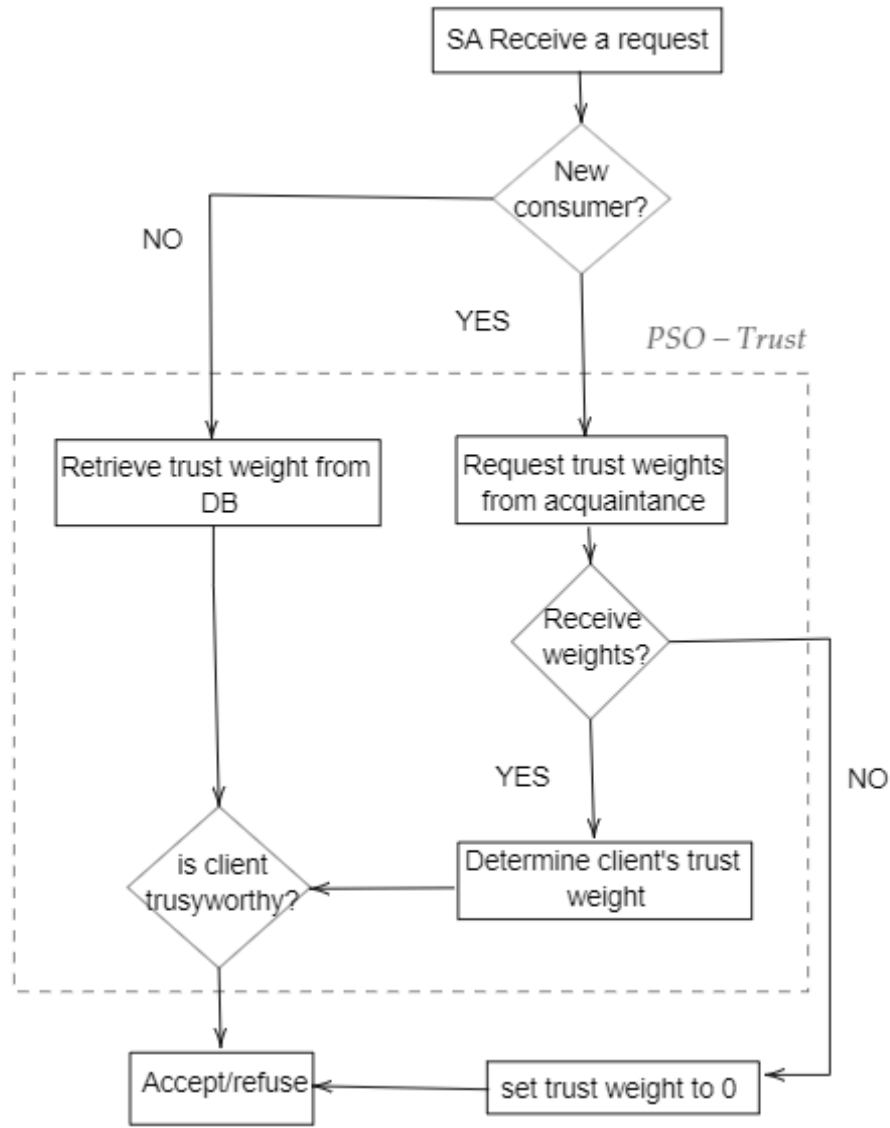

Figure 3. Steps to evaluating client's trustworthiness

If none of the acquaintances possesses information about the client, its trust weight is set to zero (as a neutral value). Otherwise, the PSO model is used to determine clients' trust weight, and is judged upon the resulting value whether to be trusted or not.

Trust values fall between [-1,1] instead of [0,1] since clients could be untrustworthy. (-1) denotes an untrustworthy agent and (1) denotes a trusty one. The median value (zero) is mostly used for agents with no history of interaction whatsoever.

### 4.3. Model presentation

The service agents (SA) are responsible for negotiating with the client and offering the needed services. When a SA is contacted by a new client C, and to ensure security, and to assign a trust weight for the new costumer, the SA would inquire its acquaintance send their trust weights of the client C. If present, the agent would receive several different weights and it would be hard to pick one from the large choice.

To overcome that, Particle Swarm Optimization is used to select the most appropriate weight in a 2D plan considering each particle as a tuple ($T_{a(x)}$, $R_{Ta(x)}$) where: $T_{A(x)}$: The trust weight of the agent A(x), and $R_{TA(x)}$: The trust value received from agent A(x).

The error of each particle is determined using Euclidian distance as a fitness function where the distance between each particle and the global best is calculated.

$$Fintess\ function = e = \sqrt{(x_1 - x_2)^2 + (y_1 - y_2)^2} \qquad (3)$$

Where: ($x_1$, $y_1$) are the particle's position and ($x_2$, $y_2$) the global best postion. The personnel best of the particles is updated if the error is minimal.

The algorithm outputs the global best of the population by generating the mean of all particles.

$$Global\ Best = \sum_{1}^{size} \frac{ParcticlePosition(i)}{size} \qquad (4)$$

Equation (4) assigns the same importance to all trust values nevertheless the positive ones should have more impact in the equation. In order to respond to that the equation is changed so that the particle position is multiplied by a variant as an impact factor (Θ) depending on the trust values:

$$Global\ Best = \frac{1}{size}\sum_{1}^{size} \Theta * ParcticlePosition(i) \qquad (5)$$

Since the population positions are between [-1,1] (to correspond to the trust weights), and so that the generated solution can be used immediately, the results of equation (5) are normalized in$[-1, 1]$.

$$Global\ Best' = 2 * \frac{GlobalBest - \min x}{\max x - \min x} - 1 \qquad (6)$$

With: x = Particle Position.

### 5. EVALUATION

First, to have a realistic view of the proposed model we started by implementing the framework in Netlogo [27].

Netlogo is a MAS modelling environment for simulating social and complex phenomenal in a world allowing the user to observe its state through time. It is mostly used by research to project the ideas into an interactive environment.

Because the cloud is open, we configure the number of agents in the population as variable, and could be increased or decreased by the user (increases even up to 150).

To have a better view of the world we chose a small population number for the illustrated experiment. At initialization (Figure 4), a random number of clients (in green) and providers (in blue) are present in the network. The "go" function allow the system to simulate the cloud environment interactions.

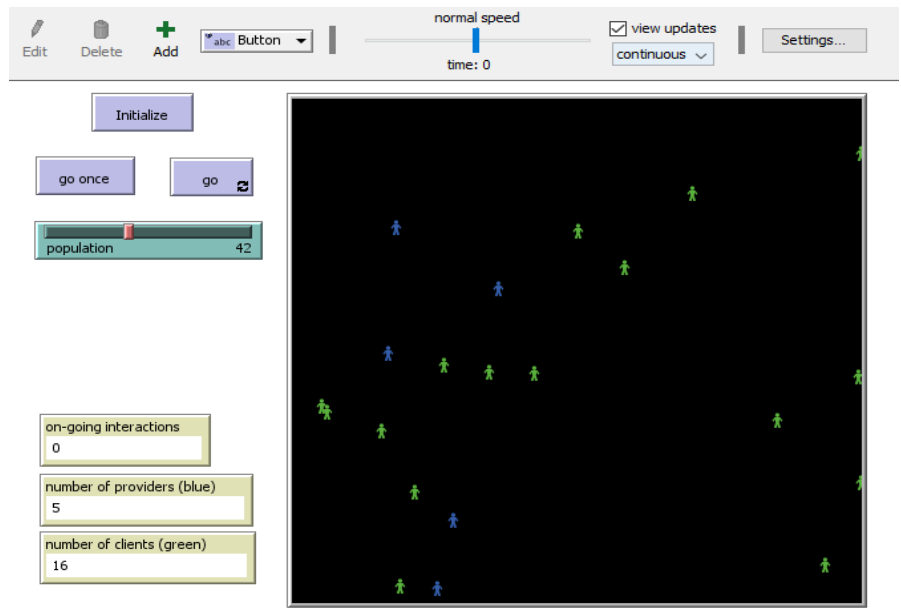

Figure 4. The Netlogo world after initialization

After running the program for some time (Figure 5), it can be notice how some providers need to contact their acquaintance after receiving requests from consumers, while others do not express this need. This is due to the presence or absence of data of the contacting clients in the providers' knowledge database.

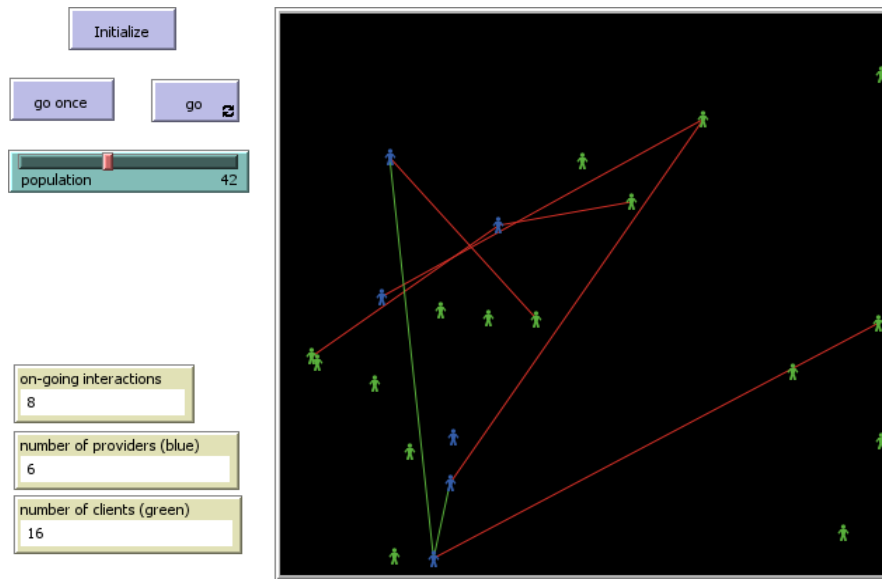

Figure 5. Ongoing interactions in the cloud

The number of agents in the cloud can change during runtime due to the openness of the cloud (agents can enter or leave the environment randomly), but when the maximum number of agents in the population is reached no more agent can enter until others leave (since the population number is previously set by the user).

To experiment our work and evaluate the performance of the PSO-Trust model, we developed it under Python.

Python offers a huge number of libraries, frameworks, and even plotting library (matplolip) that ease creating animated and interactive visualisations. Due to the lack of clients' data (for the purpose of confidentiality), we used a random dataset to evaluate the model.

The population size may differentiate depending of the received trust weights and the number of agent's acquaintance. To test the algorithm efficiency, we used different population sizes in the experiments (between 5 and 100).

To illustrate experiment's results, we start by exploiting large population; the initial high sized population present in Figure 6 (where each particle is represented by a different colour each to ease observation) is able converges after a minimal number of iterations

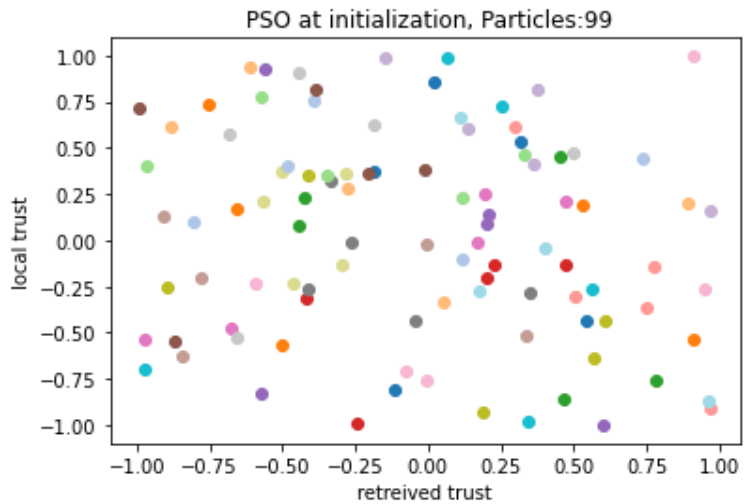

Figure 6. Particles' distribution with large population

Before conducting the experiments we stated by configuring the parameters from equation 1 and 2. While c1 represent how much confidence the particle have in itself (the local position), c2 represents how much confidence it have in its neighbourhood. Hence, since we need the particles to be more attracted to the global than personal best and converge fast, envy value is set slightly bigger than nostalgia (c2 >> c1).

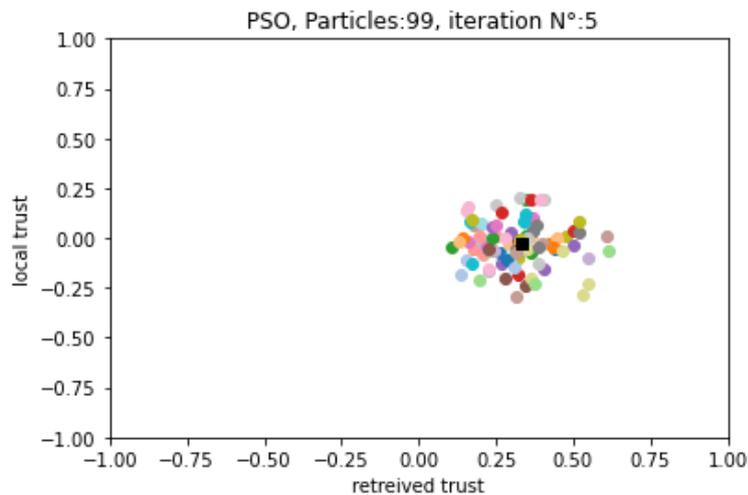

Figure 7. Convergence after five iterations.

The algorithm runs for a several number of iterations until the global best is reached or the error rate is minimum.

In Figure 7, after several iteration (five to be precise) particles have almost fully converge to the global best (represented by the black square).

Experiments also showed that even with a small swarm size (Figure 8).

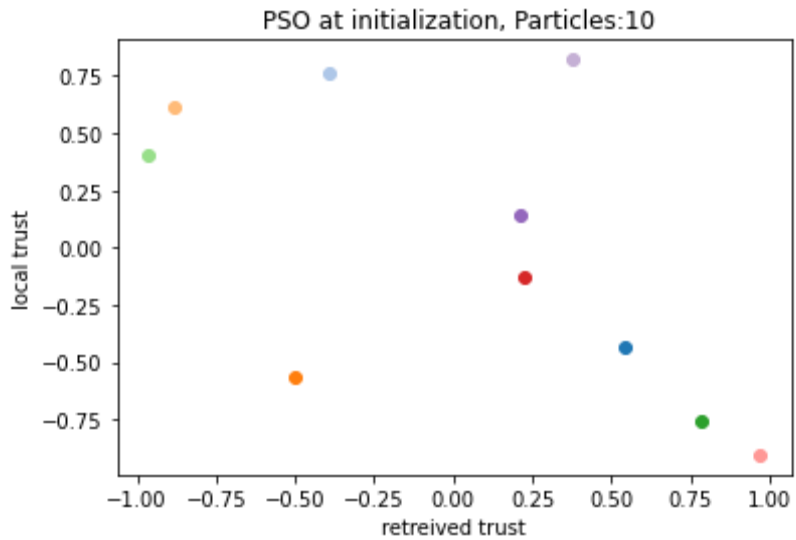

Figure 8. Small swarm size.

The algorithm succeeds this time too in finding the best trust weight value from the received values in Figure 9 (the black triangle in figure 9 represents the global best position).

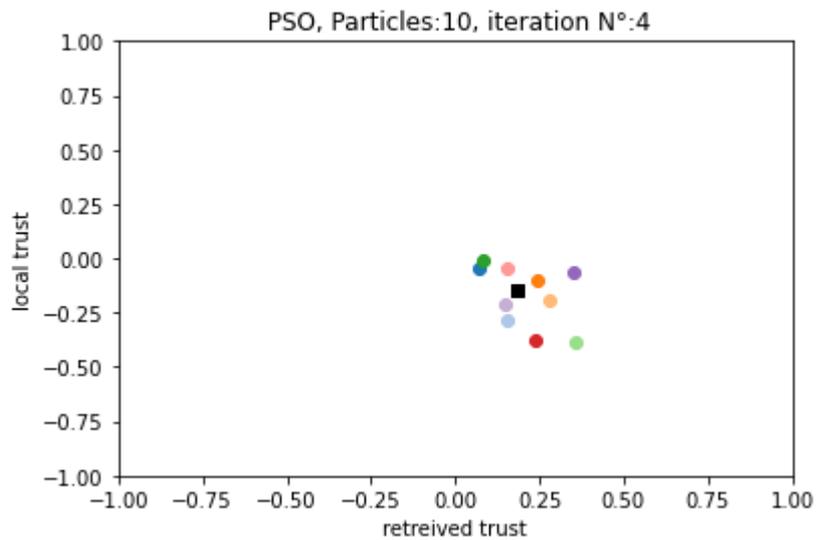

Figure 9. Swarm population near convergence

Therefore, even if the acquaintance offering information are numbered, the algorithm can still provide an accurate solution.

Because providers refuse to share their data (due to security concerns), we could not find data as to compare our proposed solution with other models and the generated data would not fit with the models description.

## 6. CONCLUSIONS AND FUTURE WORKS

Since it is the providers' responsibility to keep the data and clients secure, trust in the cloud should not be only considered by clients. Malicious users can attack the cloud and effect the data it stores, therefore, effect many clients and risk their privacy.

To secure the network and identify the clients that are worthy of trust, we proposed a PSO-Trust model where we make use of the interaction mechanisms of MAS to share reliable data.

One of the difficulties we faced is the lack of data, for this we aim to generate a database that represents clients' history and feedback from providers. For the next step, we would extend our work to construct an opponent interaction model. The current work would represent a pre-interaction phase to the interaction model and adapt the ongoing interactions based on the previous shared knowledge.